\documentclass[12pt,aps,prb,preprint]{revtex4}   % style for Physical Review B and AJP are similar

\usepackage{amsmath}    % need for subequations
\usepackage{amsfonts}  %note how statements can be commented out
\usepackage{amssymb}
\usepackage{graphicx}   % for figures

\usepackage[latin1]{inputenc}     % pour les accents
\usepackage[T1]{fontenc}

\usepackage{float}   % pour que les images soient inclusent dans le texte

\usepackage{color}   % pour faire des surlignes de couleur
\usepackage{soul}    % pour faire des surlignes de couleur

\usepackage[a4paper]{geometry} % pour récupérer les marges
\graphicspath{{./images/}}
\newtheorem{Theoreme}{Theorem}%{THEOREM}
\newtheorem{Definition}{Definition}

\newtheorem{lemma}[Theoreme]{Lemma}

\newtheorem{Remarque}{Remark}

\definecolor{lightblue}{rgb}{0.8,0.9,1} % bleu ciel

\draft

\begin{document}

\title{A heuristic extended particle 2D-model compatible with quantum mechanics}
\author{Michel Gondran}
 \affiliation{University
Paris Dauphine, Lamsade, 75 016 Paris, France}
 \email{michel.gondran@polytechnique.org}   %optional

 \author{Alexandre Gondran}
 \affiliation{Ecole Nationale de l'Aviation Civile, 31000 Toulouse, France}
  \email{alexandre.gondran@enac.fr}

\begin{abstract}
In this paper we propose an extended particle model whose evolution is deterministic. In dimension 2, the extended particle is represented by four points that define a small elastic string that vibrates, alternating between a creation process and an annihilation process. 

First we show how the spin and the Heisenberg uncertainty relations emerge from this extended particle. We then show how the complex action associated with this extended particle satisfies, from a generalized principle of least action, a second order complex  Hamilton-Jacobi equation. Third, we show that the wave function, which admits this action as a complex phase, satisfies the Schrödinger equation. Finally, we show that the gravity center of this extended particle follows the trajectories proposed by the de Broglie-Bohm interpretation well as the Schrödinger interpretation.

This model is built on two new mathematical concepts we have introduced: complex analytical mechanics on complex-valued functions and a periodic deterministic process.

\end{abstract}

\maketitle

\section{Introduction}

One of the fundamental reasons for the impossibility of synthesis between quantum mechanics and general relativity is that quantum mechanics is considered non-deterministic while relativity is considered deterministic. Most current approaches to synthesis research, such as string theory \cite{Witten} or loop quantum gravity theory\cite{Smolin, Rovelli} are based on non-deterministic general relativity. The alternative approach is to render quantum mechanics deterministic.

In this paper, we propose an extended model particle whose evolution is deterministic. In dimension 2, the extended particle is represented by four points that define the structure of a small elastic string that vibrates, alternating between a process of creation and a process of annihilation. We then show how the spin and the Heisenberg uncertainty relations emerge from this extended particle.
   
We subsequently demonstrate how the complex action associated with this extended particle satisfies, from a generalized principle of least action, a second order complex Hamilton-Jacobi equation.
We show that the wave function, which admits this action as a complex phase, satisfies the Schrödinger equation. Finally, we show that the gravity center of this extended particle follows the trajectories proposed by the de Broglie-Bohm interpretation as well the  Schrödinger interpretation.

This model is built on two new mathematical concepts: complex analytical mechanics on complex-valued functions, which we have introduced \cite{Gondran_1999, Gondran_2001a, Gondran_2001c, GondranHoblos_2003},
and the periodic deterministic process, which
we have developed\cite{Gondran_2001b, Gondran_2004}.

\bigskip

\section{An extended particle model and its trajectory}

In an orthonormal space $ \mathbb{R}^2 $, let us consider the four vertices of the unit square $ u^1=\binom{1}
{1}$, $ u^2=\binom{1}
{-1}$, $ u^3=\binom{-1}
{-1}$ and $ u^4=\binom{-1}
{1}$. There are two circular permutations of these four vertices, one $s^+$ in a clockwise direction, the other $s^-$ in the opposite direction. For each of these permutations $s\in S$ and for all $u^j$, we have $s^4 u^j= u^j$. 

We consider an extended particle represented by four points. For each time step $\varepsilon >0$
and for each of the two permutations  $s\in S$, the evolution of the four points at time  $t=n \varepsilon $ with $n=4q+r$ ( n,q,r  integer and
$0\leq r\leq 3 $), is defined by the real part of the four following discrete $Z_{\varepsilon }^{j}(t)\in \mathbb{C}^{2}$ processes:
\begin{equation}\label{eqprocessus1}
Z_{\epsilon }^{j}(n \epsilon )=Z_{\epsilon
}^{j}\left( n\epsilon -\epsilon \right)+\mathcal{V}(4q\epsilon
)\epsilon +\gamma(s^{n}u^{j}-s^{n-1}u^{j})
\end{equation}
\begin{equation}\label{eqprocessus2}
Z_{\epsilon }^{j}(0)=Z_{0}~~~~
 for~all~j,
\end{equation}
where
\begin{equation*}
\gamma= (1+i)\sqrt{\frac{%
\hslash \varepsilon }{4m}},
\end{equation*}
$\mathcal{V}(t)$\ corresponds to a continuous complex function, $\hslash $\ is the  Planck constant, $m$ the mass of the particle, and $Z_{0}$ is a given vector of
$\mathbb{C}^{2}.$

Let $\widetilde{Z}_{\epsilon }(t)$ be the solution in $\mathbb{C}%
^{2} $ of the discret system deined at time $%
t=n$ $\epsilon $ with $n=4q+r$ $\left( n\text{, }q\text{ and
}r\text{ integers and }0\leq r\leq 3\right) $ by:
\begin{equation}
\widetilde{Z}_{\epsilon }( n \epsilon
)=\widetilde{Z} _{\epsilon }(n \epsilon- \epsilon
)+\mathcal{V}(4q\epsilon )\epsilon
\end{equation}
\begin{equation}
\widetilde{Z_{\epsilon }}(0)=Z_{0}.
\end{equation}
We then verify that we have at each time
$t= n \epsilon$:
\begin{equation}\label{eqreccurenceprocess}
Z_{\epsilon }^{j}(n\epsilon )=\widetilde{Z_{\varepsilon
}}(n\epsilon )+(1+i) \sqrt{\frac{\hslash \varepsilon }{4m}}\left(
s^{n}u^{j}-u^{j}\right) .
\end{equation}

As $s^{4}u^{j}=u^{j}$, we deduce from
(\ref{eqreccurenceprocess}) that
$Z_{\epsilon }^{j}( 4q\epsilon )=\widetilde{Z}_{\epsilon }(4q
\epsilon )$~~~~for~all~j.

As $\sum^{j=4}_{j=1}s^n u^j =0$, we deduce from (\ref{eqprocessus1}) that the process $\widetilde{Z}_{\epsilon }(t)$ is the average of the four processes $Z_{\epsilon }^{j}(t)$. Its real part $\widetilde{X}_{\epsilon }(t) $  can be interpreted as the gravity center of the particle. The position $X_{\epsilon }^{j}(t)$ of each vertex j, the real part of process
$Z_{\epsilon }^{j}(t)$, satisfies the equation:
\begin{equation}\label{eqreccurenceprocessreel}
X_{\epsilon }^{j}(n\epsilon )=\widetilde{X_{\varepsilon
}}(n\epsilon )+ \sqrt{\frac{\hslash \varepsilon }{4m}}\left(
s^{n}u^{j}-u^{j}\right).
\end{equation}

\begin{figure}[H]
\begin{center}
\includegraphics[width=1.0\linewidth]{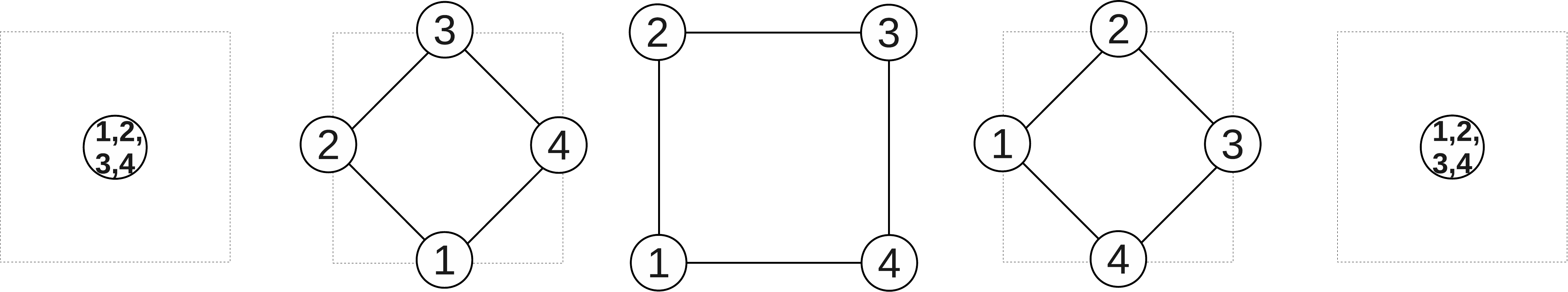}
\caption{\label{fig:sixprocessus}Evolution of the four points of the extended particle  over a period of 4 $\epsilon$ from left to right: points at  $4q\epsilon$, in extension at $( 4q+1)\epsilon$ and $(4q+2)\epsilon$, then in contraction at $(4q+3)\epsilon $ and $4(q+1)\epsilon$.}
\end{center}
\end{figure}

This equation expresses the evolution of the four points of the extended particle in relation to its center of gravity. The evolution of this extended particle over a period of 4 $\epsilon $ is shown in Figure
\ref{fig:sixprocessus}.
We consider that the four points $X_{\epsilon }^{j}(t)$
of the particle define the a string structure. The movement of the four
points corresponds to the vibration of the string.
At time $t=4 q \epsilon $, the four
points are at the center of a square and the length of the string at this time is zero. At times $t \neq 4 q \epsilon $,
it takes an extension. At times $(4q+1)\epsilon$ and
$(4 q+3)\epsilon $, the four points are at the center of the edges of
square. At time $(4q+2)\epsilon$, the four
points are on the four
vertices of the square. Moreover,
this interpretation suggests a creation process
between times $t=4 q \epsilon $ and  $(4q+1)\epsilon $ follows
an annihilation process between times $(4 q+3)\epsilon
$ and $(4q+4)\epsilon $.

The equation (\ref{eqreccurenceprocess}) leads to $
Z_{\epsilon }^{j}( t)=\widetilde{Z}_{\epsilon }(t )+
0(\sqrt{\epsilon})$ for all j and all t = n$\epsilon$.

Let $\widetilde{Z}(t)$ be the solution of the classical differential equation:
\begin{equation}
\frac{d\widetilde{Z}(t)}{dt}=\mathcal{V}(t)
\end{equation}
\begin{equation}
\widetilde{Z}(0)=Z_{0}.
\end{equation}

Because $\mathcal{V}(t)$ is continuously differentiable, we
obtain $\widetilde{Z}_{\epsilon }(t )=\widetilde{Z}(t)+
0(\epsilon)$ for all $t=n \epsilon$, and then $ Z_{\epsilon
}^{j}( t)=\widetilde{Z}(t )+ 0(\sqrt{\epsilon})$. We can deduce
\begin{Theoreme}\label{th:Convergencedesprocessus}- Each process $\widetilde{Z}^j_{\epsilon }(t )$
continuously converges towards the classical trajectory
$\widetilde{Z}(t)$ when $\epsilon\longrightarrow 0^+$.
\end{Theoreme}

Let us conclude this section with some remarks on the $Z_{\epsilon }^{j}(t) $ processes. 

\begin{Remarque}\label{r:interpretationprocessus2}- The four $Z_{\varepsilon }^{j}(t)$ processes looks like to the Nelson stochastic process \cite{Nelson_1966,Nelson_1985} based on the Wiener process, but unlike the Nelson process they are
deterministic.
However, despite being deterministic, these processes appear random with respect to the spatial extension of the string because, at time t, the rest modulo
4 of the number  $n=$ $\frac{t}{\varepsilon }$ is a
pseudorandom number.
\end{Remarque}

\begin{Remarque}\label{r:interpretationFeynman}- 
In \cite{FeynmannHibbs}, Feynmann et Hibbs
show that the "important paths" of quantum mechanics, are very irregular and nowhere
differentiable. They admit an average velocity $lim_{\Delta t\longrightarrow 0^+}\langle \frac{x_{k+1}-x_k}{\Delta
t} \rangle =v$,
but not an average quadratic velocity
because $
\langle (\frac{x_{k+1}-x_k}{\Delta t})^2 \rangle =\frac{i \hbar}{m
\Delta t}$.
The four $X_{\varepsilon }^{j}(t)$ processes satisfy the same properties as the Feynmann paths, becoming increasingly irregular and non-differentiabe when $\epsilon
\rightarrow 0^{+}$; however, the value of $\varepsilon=\Delta t$, although very
small, remains finite. 
\end{Remarque}

\section{Emergence of the spin and Heisenberg uncertainty relations}

We have associated with an extended particle four points and one cycle of four instants during the period $T =4 \epsilon$. We will assume that the properties of such a particle are the average of the properties of the four points taken on the four instants of the period.

We therefore define the average angular momentum of the extended particle satisfying
Equation(\ref{eqreccurenceprocessreel}) by:
\begin{equation*}
\sigma =E_{n,j}(\sigma^j_n)=\frac{1}{16}
\sum_{n=4q}^{4q+3}\sum_{j=1}^{4} \sigma^j_n
\end{equation*}
with
\begin{equation*}
\sigma^j_n=r^j_n \wedge p^j_n,~~~~r^j_n=X^j_\varepsilon
(n\varepsilon)= \widetilde{r}_n + \sqrt{\frac{\hbar
\varepsilon}{4m}}(s^n u^j - u^j)
\end{equation*}
and
\begin{equation*}
p^j_n=m \frac{r^j_{n+1}-r^j_n}{\varepsilon}= m
\frac{\widetilde{r}_{n+1}-\widetilde{r}_n}{\varepsilon}+
\sqrt{\frac{\hbar m}{4 \varepsilon}}(s^{n+1} u^j - s^n u^j).
\end{equation*}

By using the identity $\sum_{j=1}^{4} s^n u^j=0$ for all n,
we obtain $\sigma =\widetilde{r}\wedge m\widetilde{v}+\frac{1}{16}\hslash
\sum_{j=1}^4 (u^j \wedge s u^j)$
with $ \widetilde{r}=\frac{1}{4} \sum_{n=4q}^{4q+3} \widetilde{r}_n$
and $\widetilde{v}=\widetilde{v}(4q\varepsilon)$.
For $s=s^+$, we obtain
\begin{equation*}
\sigma \equiv \sigma _{z}=m\left( x\widetilde{v}
_{y}-y\widetilde{v}_{x}\right) -\frac{\hslash }{2}.
\end{equation*}
We deduce the theorem:
\begin{Theoreme}\label{th:emergencespin}- For all $\epsilon >0$, the extended particle which corresponds to the real part of process (\ref{eqprocessus1})(\ref{eqprocessus2}), has an average intrinsic angular momentum
$s_{z}=-\frac{\hslash }{2}$ for the permutation $s^+$ and $s_{z}=+ \frac{\hslash }{2}$ for the permutation $s^-$. 
\end{Theoreme}

\bigskip

Let $\widetilde{x}_{\varepsilon }(n\varepsilon )$ be the average position of the particle along the $x$ axis  at time
$n\varepsilon $ (with $n=4q+r$) and $m\widetilde{v}\left( 4q
\varepsilon \right) $ the average momentum. The calculation of standard deviations
$\triangle x$\ and $\Delta p_{x}$ of the position and of the momentum
along the $x,$ axis is obtained from the following equations:
\begin{equation*}
\langle \Delta x\rangle^2=\frac{1}{16}\sum_{n=4q}^{4q+3}\sum_{j=1}^{4}\left( r^j_n- \widetilde{r}_n\right) _{x}^{2}
\end{equation*}
\begin{equation*}
\langle \Delta p_{x}\rangle^2=\frac{1}{16}\sum_{n=4q}^{4q+3}\sum_{j=1}^{4}\left(
p_n^j- \widetilde{p}_n\right) _{x}^{2}
\end{equation*}
with $\widetilde{p}_n= m
\frac{\widetilde{r}_{n+1}-\widetilde{r}_n}{\varepsilon}$. We obtain $\langle \Delta x\rangle=\frac{\hbar
\varepsilon}{2m}$ and $\langle \Delta p_{x}\rangle= \frac{\hbar m}{ \varepsilon}$. We deduce the theorem:

\begin{Theoreme}\label{th:inegaliteheisenberg}- For all $\epsilon >0$
and for all $s$, the extended particle which corresponds to the real part of process (\ref{eqprocessus1})(\ref{eqprocessus2}), satisfies the Heisenberg uncertainty relations:
\begin{equation}
\langle \Delta x\rangle \cdot \langle \Delta
p_{x}\rangle = \frac{\hslash }{2}.
\end{equation}
\end{Theoreme}

\bigskip

Let us consider $f$ a twice differentiable application from $\mathbb{C}^{2}\times \mathbb{R}$ in $\mathbb{C}$.
We denote \textit{the complex Dynkin operator}, introduced by Nottale \cite{Nottale} under the name of "quantum covariant derivative":
\begin{equation}
D=\frac{\partial }{\partial t}+\mathcal{V}\cdot \triangledown -i\frac{%
\hslash }{2m}\triangle.
\end{equation}

\begin{lemma} - For all $\epsilon >0$ and for all $s $, the process
$Y_{\epsilon }(t)$ is defined by:
\begin{equation}
Y_{\epsilon }\left( t\right) =Ef(Z_{\epsilon }^{j}\left( t\right) ,t)=\frac{1%
}{4}\ \sum_{j}\left( f\left( Z_{\epsilon }^{j}\left( t\right)
,t\right) \right)
\end{equation}
\textit{with }$Z_{\epsilon }^{j}\left( t\right) $ defined by (\ref{eqprocessus1})and (\ref{eqprocessus2}),
satisfies for all $t= 4q\varepsilon$ (q integer):
\begin{equation}\label{eqdevsecondordre}
Y_{\epsilon }(t) -Y_{\epsilon }\left(t - \varepsilon  \right) =Df\left( \widetilde{Z}\left(t\right) ,t \right) \epsilon +0\left( \epsilon^2 \right).
\end{equation}
\end{lemma}

\textit{\textbf{Proof}}: First, we have $Y_{\epsilon }\left(
4q\epsilon \right) =f\left( \widetilde{Z}\left( 4q\epsilon \right)
,4q\epsilon \right) $. Using (\ref{eqreccurenceprocess}) and $%
\sum_{j}s^{n}u^{j}=0 $, we find for all $t=n \epsilon$
\begin{equation*}
Ef(Z_{\epsilon }^{j}\left( t\right) ,t)=f(\widetilde{Z}_{\epsilon
}(t),t)+
\frac{i\hslash }{4m}\epsilon E\left\{ \sum_{k,l}\frac{\partial ^{2}f(\widetilde{Z}_{\epsilon
}(t),t)}{%
\partial x_{k}\partial x_{l}}\left( s^{n}u^{j}-u^{j}\right) _{k}\left(
s^{n}u^{j}-u^{j}\right) _{l}\right\} +0 \left( \epsilon^2 \right).
\end{equation*}
For $n=4q-1$, $E (
s^{n}u^{j}-u^{j})_{k}(s^{n}u^{j}-u^{j})_{l}=\frac{4+4}{4}\delta_{kl}$
and the calculation of the last term of $Ef(Z_{\epsilon }^{j}\left(
t\right) ,t)$ yields $\frac{i\hslash }{4m}\epsilon
 2 \Delta f$. Then we deduce:
\begin{equation*}
Y_{\epsilon }\left( 4q \epsilon-\epsilon \right)
=f(\widetilde{Z}_{\epsilon
}(4q \epsilon-\epsilon ),4q\epsilon -\epsilon )+i\frac{\hslash }{2m}\epsilon \Delta f(%
\widetilde{Z}_{\epsilon }(4q \epsilon -\epsilon ),4q\epsilon -\epsilon )+0
\left( \epsilon^2 \right).
\end{equation*}
Hence, the development to first order of $f(\widetilde{Z}_{\epsilon }(4q \epsilon-\epsilon ),4q \epsilon-\epsilon )$ leads to equation (\ref{eqdevsecondordre}).$\Box $

\section{Second order Complex Hamilton-Jacobi equation}

We will show that the evolution of the process $Z_{\varepsilon }^{j}(t) $ defined by the equations (\ref{eqprocessus1})(\ref{eqprocessus2}), is also given by a second order complex Hamilton-Jacobi equation.  To do this, we use \textit{a complex analytical mechanics}
and \textit{a generalized principle of least action}. The complex analytical mechanics is a generalization of classic analytical mechanics
but with objects having a complex position
$Z(t)\in \mathbb{C}^{2}$ and a complex velocity $\mathcal{V}\left(
t\right) \in \mathbb{C}^{2}$. We use the complex minimum of a complex function 
and the complex Minplus analysis introduced in
\cite{Gondran_1999,Gondran_2001a,Gondran_2001c,GondranHoblos_2003}. We recall the principle in the following definitions.

\begin{Definition}- For a complex function $f\left( Z\right) =f\left( X +i Y\right)
$ from $\mathbb{C}^{n}$ in $\mathbb{C}$ such as $f\left( Z\right) =P\left( X,Y\right) +iQ\left( X,Y\right)
$, we define the complex minimun, if it exists, by
$\min \left\{ f\left( Z\right) /Z\in \mathbb{C}^{n}\right\}=f\left( Z_{0}\right)$  where $\left( X_{0},Y_{0}\right) $ is a saddle point of $P\left( X,Y\right)$:
$P\left( X_{0},Y\right) \leq P\left( X_{0},Y_{0}\right) \leq
P\left( X,Y_{0}\right)$ $\forall
(X,Y)\in R^{n}\times R^{n}$.
\end{Definition}

A complex function $f\left( Z\right) $ is (strictly) \textit{convex} if $P\left( X,Y\right) $\ is (strictly) convex in $X$\ and (strictly) concave in $Y$. 

If $f\left( Z\right)$ is a holomorphic function, then  a necessary condition for $Z_0$ to be a minimum of $f(Z)$ in $\mathbb{C}^{n} $ is $\nabla f(Z) =0$. It is sufficient if $f(Z)$ is also convex.

Using the classical Lagrange function $L(x,\dot{x},t)$, an analytical function in $x$ and $\dot{x}$, we define the complex Lagrange function $L(Z,\mathcal{V},t)$.

\begin{Definition}- With all complex and strictly convex functions $f(Z)$, we associate a complex Fenchel-Legendre transform $\widehat{f}\left( P\right)$
defined by:
\begin{equation*}
\widehat{f}\left( P\right) =\underset{Z\in \mathbb{C}^{n}}{\mathit{\max }}%
\left( P.Z-f\left( Z\right) \right)
\end{equation*}
\end{Definition}

Before presenting our generalization of the principle of least action for our extended particle model, let us recall this principle for the definition of the Hamilton-Jacobi action in the classical case.
At time 0, an initial action $S^0(\textbf{x})$, a function of
$\mathbb{R}^{n}$ in $\mathbb{R}$ is given. This initial action corresponds to the initial velocity field $\textbf{v}_0(\textbf{x})= \nabla S^0(\textbf{x})/m $.

The Hamilton-Jacobi action $S(\mathbf{x},t)$ is then the function 
\begin{equation}\label{eq:defactionHJ}
S(\mathbf{x},t)=\underset{\textbf{x}_0;\mathbf{u}\left( s\right),0 \leq
s\leq t }{\min
}\left\{ S_{0}\left( \mathbf{x}_{0}\right) +\int_{0}^{t}L(\textbf{x}(s),%
\mathbf{u}(s),s)ds\right\}
\end{equation}
where the minimum of (\ref{eq:defactionHJ}) is taken on all the trajectories 
with $\textbf{x}_0$ as position at the initial time and $\textbf{x}$ as position at time t, and on all the velocities $\mathbf{u}(s)$, $s\in$ $\left[ 0,t\right]$.

The action $S(\mathbf{x},t)$ defined by (\ref{eq:defactionHJ}) can
be decomposed into
\begin{equation*}
S(\mathbf{x},t)=\underset{\textbf{x}_0;\mathbf{u}\left( s\right),0
\leq s\leq t }{\min
}\left\{ S_{0}\left( \mathbf{x}_{0}\right) +\int_{0}^{t -dt}L(\textbf{x}(s),%
\mathbf{u}(s),s)ds +\int_{t - dt}^{t}L(\textbf{x}(s),%
\mathbf{u}(s),s)ds\right\}
\end{equation*}
and then satisfies, between the time t-dt and t, the optimality equation:

\begin{equation*}
S(\mathbf{x},t)=\underset{\mathbf{u}\left(s\right) , t-dt\leq
s\leq t}{\min }\left\{ S(\mathbf{x}- \int^{t}_{t-dt}\textbf{u}(s)
ds,t-dt )
+\int_{t-dt}^{t}L(\textbf{x}(s),\mathbf{u}(s),s)ds\right\}.
\end{equation*}

Assuming that S is differentiable in \textbf{x} and t, L
differentiable in \textbf{x}, \textbf{u} and t, and \textbf{u}(s)
continuous, this equation becomes:
\begin{equation}\label{eq:defactionlocale}
S(\mathbf{x},t)=\underset{\mathbf{u}\left(t\right)}{\min }\left\{ S(\mathbf{x}-\textbf{u}(t)
dt,t- dt) + L(\textbf{x},
\mathbf{u}(t),t)dt +\circ \left( dt\right)\right\},
\end{equation}
\begin{equation*}
0=\underset{\mathbf{u}\left( t\right) }{\min }\left\{
-\frac{\partial S}{
\partial \mathbf{x}}\left( \mathbf{x} \mathbf{,}t\right)
\textbf{u}\left( t\right) dt-\frac{\partial S}{\partial \mathbf{t%
}}\left( \mathbf{x} \mathbf{,}t\right) dt+L(\mathbf{x},
\mathbf{u}(t),t)dt+\circ \left( dt\right) \right\}
\end{equation*}
and dividing by $dt$ and letting $dt$ tend towards $0^{+}$,
\begin{equation*}
\frac{\partial S}{\partial \mathbf{t}}\left( \mathbf{x,}t\right)
=\underset{ \mathbf{u}}{\min }\left\{
L(\mathbf{x},\mathbf{u},t)-\mathbf{u\cdot }\frac{
\partial S}{\partial \mathbf{x}}\left( \mathbf{x,}t\right)
\right\}.
\end{equation*}

It is the classical Hamilton-Jacobi equation:
\begin{equation*}
\frac{\partial S}{\partial t}\left( \mathbf{x,}t\right)
+H(\mathbf{x},\frac{
\partial S}{\partial \mathbf{x}}{\LARGE ,}t)=0\text{ \ }
\end{equation*}
where\ $H(\mathbf{x},\mathbf{p},t)$\textit{\ } is the Fenchel-Legendre transform of
$L(\mathbf{x},\mathbf{u},t)$.

We can now define a complex Hamilton-Jacobi action for the extended particle which corresponds to processes (\ref{eqprocessus1})(\ref{eqprocessus2}).

At time 0, we take a complex action $\mathcal{S}^{0}\left( Z\right) $, which is a holomorfic function from $\mathbb{C}^{2}$ in $\mathbb{C}$. The complex Hamilton-Jacobi action associated with processes (\ref{eqprocessus1})(\ref{eqprocessus2}) is obtained from a generalization of the optimality equation (\ref{eq:defactionlocale}).

\begin{Definition}- The complex action
$\mathcal{S}_{\varepsilon }(Z,t)$ satisfies the following optimality equation defined at times $ t=4q\varepsilon $:
\begin{equation}\label{eq:principemoindreactiongeneral}
\mathcal{S}_{\varepsilon }(Z ,t)=\min_{\mathcal{V}%
\left( t\right) }\frac{1}{4}\sum_{j}\left\{
\mathcal{S}_{\varepsilon
}(Z- \mathcal{V}(t)\epsilon -\gamma(s^4 u^j -s^3u^j),t-\epsilon )+L(Z,\mathcal{%
V}(t),t)\epsilon \right\}
\end{equation}
where the minimum is a complex minimum on the possible complex velocities $\mathcal{V}\left( t\right)$. For
$t=0$, we have the initial condition:
\begin{equation*}
\mathcal{S}_{\epsilon }\left( Z,0\right) =\mathcal{S}^{0}\left(
Z\right) \text{ \ \ \ \ \ \ }\forall Z\in \mathbb{C}^{2}.
\end{equation*}
\end{Definition}

At time $t= 4q \epsilon$, we have $
\widetilde{Z_{\varepsilon
}}(t )= Z_{\epsilon }^{j}(t)= Z_{\epsilon }^{j}(t-\varepsilon)+\mathcal{V}(t
)\epsilon +\gamma(s^{4}u^{j}-s^{3}u^{j})$.

We obtain the optimality equation (\ref{eq:principemoindreactiongeneral}) from the optimality equation (\ref{eq:defactionlocale}) by identifying $\epsilon$ with dt, $\textbf{x}$ with $\widetilde{Z_{\varepsilon
}}(t ) $ and $\mathbf{x}-\textbf{u}(t)dt $ with $Z_{\epsilon }^{j}(t-\varepsilon) $.
The equation (\ref{eq:principemoindreactiongeneral}) can be interpreted as \textit{a new least action principle} adapted to the process defined by (\ref{eqprocessus1})(\ref{eqprocessus2}).
In this case, the decision on velocity takes place only at times $t=4q\epsilon $, i.e. at times corresponding to annihilation-creation.

\bigskip

\begin{Theoreme}\label{th:eqschrodingeraction}- If a complex process satisfies the new least action principle (\ref{eq:principemoindreactiongeneral}) and if the Lagrangian is 
$L(x,\dot{x},t)= \frac{1}{2}m\dot{x}^{2}-V\left( x\right) $, then the complex action satisfies the second order complex Hamilton-Jacobi' equation:
\begin{equation}\label{eqHJsecondordrecomplexe1}
\frac{\partial \mathcal{S}}{\partial t}+\frac{1}{2}\left(
\triangledown \mathcal{S}\right) ^{2}+V\left( Z\right)
-i\frac{\hslash }{2m}\triangle
\mathcal{S}=0\text{ \ \ \ \ \ \ }\forall \left( Z,t\right) \in \mathbb{C}%
^{2}\times \mathbb{R}^{+}
\end{equation}
\begin{equation}\label{eqHJsecondordrecomplexe2}
\mathcal{S}\left( Z,0\right) =\mathcal{S}^{0}\left( Z\right)
\text{ \ \ \ \ \ \ }\forall Z\in \mathbb{C}^{2}.
\end{equation}
\end{Theoreme}

\textbf{Proof}: We only do a formal proof
assuming $ \mathcal{S}_{\varepsilon }(Z,t)$ is a very regular function in $\varepsilon$, holomerphic in Z and differentiable in t. Lemma\ref{eqprocessus1} yields:
\begin{equation*}
\frac{1}{4}\sum_{j}\left\{ \mathcal{S}_{\epsilon
}(Z-\mathcal{V}(t)\epsilon -\gamma(s^4 u^j- s^3u^j)),t- \epsilon
)\right\} =\mathcal{S}_{\epsilon }(Z,t)-D\mathcal{S}_{\epsilon
}(Z,t)\epsilon +0(\epsilon^2).
\end{equation*}
We deduce in $\left( Z,t\right) $ the equation:

\begin{equation}\label{eqtransformedefenchel}
\frac{\partial \mathcal{S}_{\epsilon }}{\partial t}=\underset{\mathcal{V}}{%
\text{\textit{min}}}\left( L(Z,\mathcal{V}{\LARGE
,}t)-\mathcal{V}\cdot
\triangledown \mathcal{S}_{\epsilon }+i\frac{\hslash }{2m}\triangle \mathcal{%
S}_{\epsilon }+0\left( \epsilon \right) \right)
\end{equation}
hence the theorem, letting $\epsilon $ tend towards 0$%
^{+} $, and finally taking the complex Fenchel
transform of $L(Z,\mathcal{V},t)$.$\Box $

\bigskip

%Remarquons que $\mathcal{S}(Z,t)$, fonction classique d\'{e}finie
%sur $\mathbb{C}^{2}\times \mathbb{R}^{+}$ par
%(\ref{eqHJsecondordrecomplexe1})(\ref{eqHJsecondordrecomplexe2}),
%ne d\'{e}pend pas du rep\`{e}re orthonorm\'{e}.

\section{ Schrödinger equation}

Taking $\Psi(Z,t)
=e^{i\frac{\mathcal{S}(Z,t)}{\hslash }}$ as the wave function and applying the restriction of
(\ref{eqHJsecondordrecomplexe1})(\ref{eqHJsecondordrecomplexe2})
to the real part of Z, theorem \ref{th:eqschrodingeraction} becomes:

\begin{Theoreme}\label{th:schrodingerreel}- If the complex process satisfies the new least action principle (\ref{eq:principemoindreactiongeneral}) and if the Lagrangian is $L(x,\dot{x},t)=
\frac{1}{2}m\dot{x}^{2}-V\left( x\right) $, then the wave function $\Psi $ satisfies the Schrödinger equation:
\begin{equation*}
i\hslash \frac{\partial \Psi }{\partial t}=\mathcal{-}\frac{\hslash ^{2}}{2m}%
\triangle \Psi +V(X)\Psi \qquad \forall (X,t)\in
\mathbb{R}^{2}\times \mathbb{R}^{+}
\end{equation*}
\begin{equation*}
\Psi (X,0)=\Psi ^{0}(X)\qquad \forall X\in \mathbb{R}^{2}.
\end{equation*}
\end{Theoreme}

As $L(Z,\mathcal{V},t)=\frac{1}{2}m\mathcal{V}^{2}-V\left(
Z\right) $, the minimum of (\ref{eqtransformedefenchel}) is obtained with $m\mathcal{V-}\triangledown \mathcal{S}_{\epsilon }=0$. Then, we have
\begin{equation}\label{vitesseimagine}
\mathcal{V}\left( t\right) =\frac{\nabla \mathcal{S}\left(
Z,t\right) }{m}
\end{equation}

If we take $\Psi(X,t)=\sqrt{\rho}(X,t)e^{i\frac{S(X,t)}{\hslash}}$ and by breaking down $\mathcal{S}(Z,t)$\ into its real and imaginary parts,
\begin{equation}\label{vitesseimagine2}
\mathcal{S}(X,t)=S(X,t)-ilog\rho(X,t)\frac{\hslash}{2},
\end{equation}
we can deduce that the the trajectory $\widetilde{X}(t)$ of the center of gravity 
satisfies the classical differential equation:
\begin{equation}\label{eqBBtrajectoire}
\frac{d\widetilde{{\large X}}(t)}{dt}=\frac{\nabla S}{m},\text{ \
\ \ \ \ \ \ \ \ }\widetilde{{\large X}}(0)={\large X}_{0}.
\end{equation}
This is the trajectory proposed by de Broglie
\cite{Broglie_1927} and Bohm \cite{Bohm_1952}.

\begin{Theoreme}\label{th:schrodingerreel}- If a complex process satisfies the new least action principle (\ref{eq:principemoindreactiongeneral}) and if the Lagrangian is $L(x,\dot{x},t)=
\frac{1}{2}m\dot{x}^{2}-V\left( x\right) $, then the real part of the gravity center follows the trajectory proposed by Broglie and Bohm.
\end{Theoreme}

A fundamental property of this trajectory is that the density of probability $\varrho (x,t)$\ of a family of particles satisfying (\ref{eqBBtrajectoire}) and having a probability density $\rho _{0}\left(
x\right) $ at initial time, satisfies the Madelung continuity equation:
\begin{equation*}
\frac{\partial \varrho }{\partial t}+div(\varrho \frac{\nabla
S}{m})=0 \label{Math25}
\end{equation*}
so that the trajectories are consistent with the Copenhagen interpretation.

\begin{Remarque}\label{r:surepsilon}- To specify the model, we must make a choice of $\varepsilon $. The most natural hypothesis is to link it to the de Broglie wavelength or  to the Compton wavelength. Now, the internal motion of the process defined by
(\ref{eqprocessus1})(\ref{eqprocessus2}) has a period of $
4\varepsilon .$ We can identify this period with the de Broglie frequency
and put
\begin{equation}
4\varepsilon =T=\frac{\lambda _{dB}}{v}=\frac{h}{mv^{2}}
\end{equation}
or with the Compton frequency and put
\begin{equation}
4\varepsilon =T=\frac{\lambda _{C}}{c}=\frac{h}{mc^{2}}.
\end{equation}
With the hypothesis of the de Broglie  wavelength, $\varepsilon $ varies along the trajectory as a function of the velocity of the particle. With the assumption of the Compton wavelength, $\varepsilon $ stays constant along the path.
\end{Remarque}

\begin{Remarque}\label{r:surepsilon}- Let us answer the question about the meaning of the imaginary speed (\ref{vitesseimagine}) for our particle model in real space $\mathbb{R}^{2}$. It is possible to show\cite{Holland,Gondran_2003} that for particles with a constant spin $\textbf{s}$, the Dirac equation implies that the momentum of a particle is given by
\begin{equation}\label{vitessespin}
m\mathcal{V}=\nabla S+ \nabla log\rho\times \textbf{s}
\end{equation}
where the second term corresponds to the spin-dependent current (Gordon current).
In dimension 2, we have $\textbf{s}=\frac{\hslash}{2}\sigma_x \sigma_y$. The bivector $\sigma_x \sigma_y$ in the Clifford algebra $Cl_2$ is represented by the imaginary number i,  and the velocity (\ref{vitessespin}) is the gradient of the quaternion $ S+ log\rho \textbf{s}$ as $\nabla S(X,t) -i \nabla log\rho(X,t) \frac{\hslash}{2}$ is the gradient of the complex function (\ref{vitesseimagine2}).
\end{Remarque}

\section{Conclusion}
We presented a particle model in dimension 2 which seems compatible with quantum mechanics, especially with Louis de Broglie's particle wave duality\cite{Broglie_1927}.
One possible interpretation of the imaginary part of the process
$Z_{\epsilon }^{j}(t)$ corresponds to the bivector $\sigma_x \sigma_y $ in the Clifford algebra $Cl_2$. The particle model in dimension 3 will involve a variable spin based on the Clifford algebra $Cl_3$ or $Cl(1,3)$ obtaining the Pauli or Dirac equation instead of the Schrödinger equation.

\end{document}